\documentclass[12pt,pdftex,a4paper,twoside]{scrartcl}
\pdfoutput=1
\usepackage[ngerman,english]{babel}
\usepackage[utf8]{inputenc}
\usepackage[T1]{fontenc}
\usepackage[pdftex]{graphicx}
\usepackage{caption}
\usepackage{subcaption}
\usepackage{braket}
\usepackage[strings]{underscore}
\usepackage{slashed}
\usepackage{lmodern}
\usepackage{xcolor}
\usepackage{pifont}
\newcommand{\cmark}{\ding{51}}
\newcommand{\xmark}{\ding{55}}
\usepackage{physics}
\usepackage{color}
\usepackage{isotope}
\usepackage{cite}
\usepackage{braket}
\usepackage{units}
\usepackage{siunitx}
\usepackage{nicefrac}
\usepackage{xfrac}
\usepackage{thumbpdf}
\definecolor{webgreen}{rgb}{0,.5,0}
\usepackage[pdftex,colorlinks=true,urlcolor=blue,linkcolor=black,citecolor=black]{hyperref}
\usepackage[]{marvosym}
\usepackage[]{textcomp}
\usepackage[]{mathcomp}
\usepackage{lscape}
\usepackage{pdflscape}
\usepackage[]{amsmath,amsfonts,amssymb,wasysym}
\usepackage{bm}
\usepackage{booktabs}
\usepackage{array}
\usepackage{float}
\usepackage{tabularx}
\usepackage{lipsum}
\usepackage{listings}
\graphicspath{{./pictures/}} 
\makeatletter
\DeclareOldFontCommand{\rm}{\normalfont\rmfamily}{\mathrm}
\DeclareOldFontCommand{\sf}{\normalfont\sffamily}{\mathsf}
\DeclareOldFontCommand{\tt}{\normalfont\ttfamily}{\mathtt}
\DeclareOldFontCommand{\bf}{\normalfont\bfseries}{\mathbf}
\DeclareOldFontCommand{\it}{\normalfont\itshape}{\mathit}
\DeclareOldFontCommand{\sl}{\normalfont\slshape}{\@nomath\sl}
\DeclareOldFontCommand{\sc}{\normalfont\scshape}{\@nomath\sc}
\makeatother

\usepackage[a4paper,text={150mm,235mm},centering,twoside,bindingoffset=5mm]{geometry}
\usepackage{fancyhdr}
\fancyheadoffset{8mm}
\fancyfootoffset{8mm}
\newcommand{\pagenumfont}{\fontfamily{cmor}\selectfont}
\newcommand{\mypagenumber}{\begin{tabular}{p{10mm}}\hline\parbox[][][c]{10mm}{\centering\vspace*{2mm}\pagenumfont\thepage}\end{tabular}}
\fancypagestyle{plain}{
 \fancyhf{}
  \fancyfoot[R]{\mypagenumber}

}

\begin{document}
\renewcommand{\baselinestretch}{1.0}

\pagenumbering{arabic}
\title{\LARGE{Distinguishing Axion Models with IAXO}}
\date{}
\author{\Large{Joerg Jaeckel and Lennert J. Thormaehlen}  \\[0.2cm] \small{Institut f\"ur Theoretische Physik, Universit\"at Heidelberg,}\\[-0.2cm]\small{Philosophenweg 16, 69120 Heidelberg, Germany}}
\pagestyle{plain}
\maketitle

\begin{abstract}
\vspace*{-1.0cm}
Helioscopes, such as the proposed International Axion Observatory (IAXO), have significant discovery potential for axions and axion-like particles. In this note, we argue that beyond discovery they can resolve details of the model. In particular, in the region suggested by stellar cooling anomalies, there is a good chance to measure the mass of the particle and separately its couplings to electrons and photons. This can give crucial information on the nature of the underlying model. 
To achieve this, energy resolved detectors and a setup with low energy threshold are needed.
\end{abstract}

\section{Introduction}
The International Axion Observatory (IAXO)~\cite{Irastorza:2011gs,Armengaud:2014gea,Irastorza:2013dav} is a next generation axion helioscope~\cite{Sikivie:1983ip} and, when finished, will be the most sensitive broadband search for solar axions so far. It will explore new parameter space including KSVZ~\cite{Kim:1979if,Shifman:1979if} and DFSZ~\cite{Zhitnitsky:1980tq,Dine:1981rt} type axions -- motivated by the Peccei-Quinn solution to the strong CP-problem~\cite{Peccei:1977hh, Peccei:1977ur,Weinberg:1977ma,Wilczek:1977pj} -- in the mass range of \SIrange[range-units = brackets]{1}{100}{\milli \electronvolt} \cite{Irastorza:2018dyq}. More recently, interesting hints for axions and axion-like particles have arisen from stellar cooling anomalies~\cite{Isern:1992gia,Isern:2008nt,Corsico:2012ki,Corsico:2012sh,Viaux:2013lha,Corsico:2014mpa,Ayala:2014pea,Bertolami:2014wua,Bertolami:2014noa,Straniero:2015nvc,Arceo-Diaz:2015pva,Giannotti:2015kwo,Corsico:2016okh,Battich:2016htm} (for a combined discussion see~\cite{Giannotti:2017hny}) and the transparency of the universe to high-energy $\gamma$-rays \cite{DeAngelis:2007dqd,Mirizzi:2007hr,Simet:2007sa,DeAngelis:2008sk,SanchezConde:2009wu,Dominguez:2011xy,DeAngelis:2011id,Essey:2011wv,Horns:2012fx,Meyer:2013pny,Rubtsov:2014uga,Kohri:2017ljt,Korochkin:2018tll}.\footnote{See~\cite{Sanchez:2013lla,Dominguez:2015ama,Biteau:2015xpa} for analyses that do not find hints in this direction.} Additional motivation arises from the possibility of such particles being dark matter~\cite{Preskill:1982cy, Abbott:1982af,Dine:1982ah, Turner:1985si,Arias:2012az}.

The solar axion spectrum depends on the properties of the axion or more specifically on how strongly it couples to photons and electrons. In this note, we make use of the difference in the resulting axion/axion-like particle (ALP) spectra to determine these two couplings separately.

For our phenomenological purposes we consider an axion or ALP coupled to photons and electrons,
\begin{equation}
\label{lagrangian}
\mathcal{L}_a= \frac{1}{2}(\partial_\mu a)(\partial^\mu a)-\frac{1}{2}m_a^2a^2-\frac{g_{a\gamma}}{4}aF_{\mu \nu}\tilde{F}^{\mu \nu}+\frac{g_{ae}}{2m_e}\partial_\mu a(\bar{e}\gamma^\mu\gamma^5e).
\end{equation}

The QCD axion models all introduce a new complex scalar to the Standard Model but differ in the other (heavy) degrees of freedom. This results in different values for the electron and photon couplings, making it possible to learn something about the underlying model despite the fact that all the other extra particles are too heavy to detect.
Important examples are the KSVZ and DFSZ models~\cite{Kim:1979if,Shifman:1979if,Zhitnitsky:1980tq,Dine:1981rt}. In addition to the singlet, they introduce a new heavy quark or a second Higgs doublet, respectively. In the KSVZ model, axions do not couple to electrons at tree level. Requiring perturbativity constrains the coupling to electrons in the DFSZ~
I (leptons and down-type quarks couple to the same Higgs doublet) and DFSZ~
II (leptons and up-type quarks couple to the same Higgs doublet) models~\cite{Irastorza:2018dyq,Chen:2013kt,Giannotti:2017hny} as shown in Tab.~\ref{couplingstrengths}. In our figures we show the DFSZ model-space as an example. However, our analysis strategy is model-independent.
\begin{table}
\centering
\renewcommand{\arraystretch}{1.3}
\begin{tabular}{|l|c|c|}
	\hline
Model &$C_{a\gamma}=\frac{2\pi f_a g_{a\gamma}}{\alpha}$& $C_{ae}=\frac{f_ag_{ae}}{m_e}$ \\
\hline
KSVZ & -1.92&$\sim$\num{2e-4}\\
DFSZ I& 0.75&(0.024, 0.33)\\
DFSZ II & -1.25&(-0.31, 0)\\
\hline

\end{tabular}
\caption{Possible range of coupling strengths in the KSVZ and DFSZ models in terms of the dimensionless couplings $C_{a\gamma}$ and $C_{ae}$ from~\cite{Irastorza:2018dyq}. $\alpha$ is the fine-structure constant, $f_a$ the axion decay constant and $m_e$ the electron mass.}
\label{couplingstrengths}
\end{table}

In order to get information on the underlying fundamental model that gives rise to the axion or ALP, it is crucial to individually measure as many of its properties as possible. 
In this work, we show that a helioscope~\cite{Sikivie:1983ip} such as IAXO~\cite{Irastorza:2011gs,Armengaud:2014gea,Irastorza:2013dav} can indeed achieve more than just the discovery, but potentially allows us to measure $g_{a\gamma}$ and $g_{ae}$ separately. In addition, we consider the possibility to measure the mass in such a setup\footnote{Using a buffer gas and the spectral distortions caused by the decoherence of the axion signal as a possible avenue to a mass measurement has already been suggested in~\cite{Irastorza:2018dyq,Arik:2008mq}.}. This allows us to gain important information on the underlying model. For example, we can check whether the measured mass and couplings are consistent with the same value of the axion decay constant, thereby providing evidence whether we are dealing with a true QCD axion.\footnote{To establish a true QCD axion, it would of course be desirable to directly measure the defining gluon coupling.} In combination with information on the electron coupling, this would then point us in the direction of KSVZ or DFSZ axions.

This note is structured as follows. In Sect.~\ref{cooling}, we briefly recall the hints from stellar cooling anomalies that we use as the benchmark scenario for our study. We also review the spectrum of axions emitted by the sun
and discuss how this allows us to distinguish between the electron and the photon couplings at axion helioscopes. Sect.~\ref{simulation} describes our simulation and presents the main physics results. We summarize and conclude in Sect.~\ref{summary}.

Before we continue, let us briefly note that in the following, we talk, for brevity, only about axions. Still, our results equally apply to more general axion-like particles and, whenever we say axions, axion-like particles are meant to be included.

\section{Stellar cooling anomalies, the solar axion spectrum and its observation at helioscopes}\label{cooling}
\subsection{Stellar cooling}
The observation of white dwarfs \cite{Isern:2008nt,Bertolami:2014noa,Bertolami:2014wua,Isern:1992gia,Corsico:2012ki,Corsico:2012sh,Corsico:2014mpa,Corsico:2016okh,Battich:2016htm} as well as horizontal branch stars and red giants \cite{Ayala:2014pea,Straniero:2015nvc,Viaux:2013lha,Arceo-Diaz:2015pva} indicate the potential for additional cooling, compared to the cooling predicted by models of stellar evolution, in which only the Standard Model of particle physics is taken into account (see, e.g.\ \cite{Giannotti:2017hny} for a summary of the stellar hints).

Axions would be able to explain this deviation between observation and theory. They can be produced inside stars through a number of different production channels. Importantly, due to their weak coupling to Standard Model particles, most of them would escape immediately and contribute significantly to the cooling of stars. 
Depending on the stellar environment, the weight between the production channels via the photon ($g_{a\gamma}$) and electron ($g_{ae}$) couplings, and therefore the strength of the axion cooling, changes. Combining the different observations gives an approximate best fit point~\cite{Giannotti:2017hny},
\begin{equation}
g_{a\gamma}\approx(1.4\pm 2.5)\times 10^{-11}\,{\rm GeV}^{-1}\qquad g_{ae}\approx (1.5\pm0.5)\times10^{-13}.
\label{best_fit}
\end{equation}
More precise $1,2$ and $3$ $\sigma$ curves are given in~\cite{Giannotti:2017hny}.
Notably there is an indicative preference for a non-vanishing electron coupling, whereas the photon coupling could be small or vanishing. 

As long as the rest mass of the axion is smaller than the core temperature of the observed object, its mass does not influence the amount of additional cooling. 

In the following, we choose this best fit as our benchmark scenario and indicate the confidence contours of~\cite{Giannotti:2017hny} in our sensitivity plots.
We follow~\cite{Giannotti:2017hny} and focus on DFSZ axion models for illustration. However, our phenomenological analysis is fully specified in terms of the photon and electron couplings as well as the mass. 

Similarly, supernovae \cite{Turner:1987by,Giannotti:2017hny} and neutron stars \cite{Leinson:2014ioa} set stringent limits on an axion-nucleon coupling. Neutron stars also provide a hint towards the existence of ALPs \cite{Leinson:2014ioa,Tanabashi:2018oca}. However, Helioscopes are not sensitive to these couplings and they cannot be inferred from $g_{a\gamma}$ and $g_{ae}$ in a model-independent way. This is why we will leave these results out in our analysis even though they can provide additional constraints for specific models \cite{Giannotti:2017hny}.

\subsection{Solar axion spectrum and spectral properties of helioscopes}\label{spectrum}
Two main factors determine the photon spectrum observed inside an axion helioscope.
The first is the spectrum of axions emitted by the sun. This is most directly influenced by the size of the photon and electron coupling, respectively, and will allow us to distinguish between the two.
The second are the spectral properties of the helioscope itself,
i.e.\ the conversion probability for an axion of a given energy multiplied by the efficiency with which it arrives at the photon detector. To distinguish between the photon and electron couplings, it is important to understand how this affects the original spectrum. But, we will also see that suitable changes in these experimental features can be used to facilitate the distinction between the two couplings. We will now discuss both of these ingredients.

\bigskip

\begin{figure}[t]
\centering
\includegraphics[width=.7\linewidth]{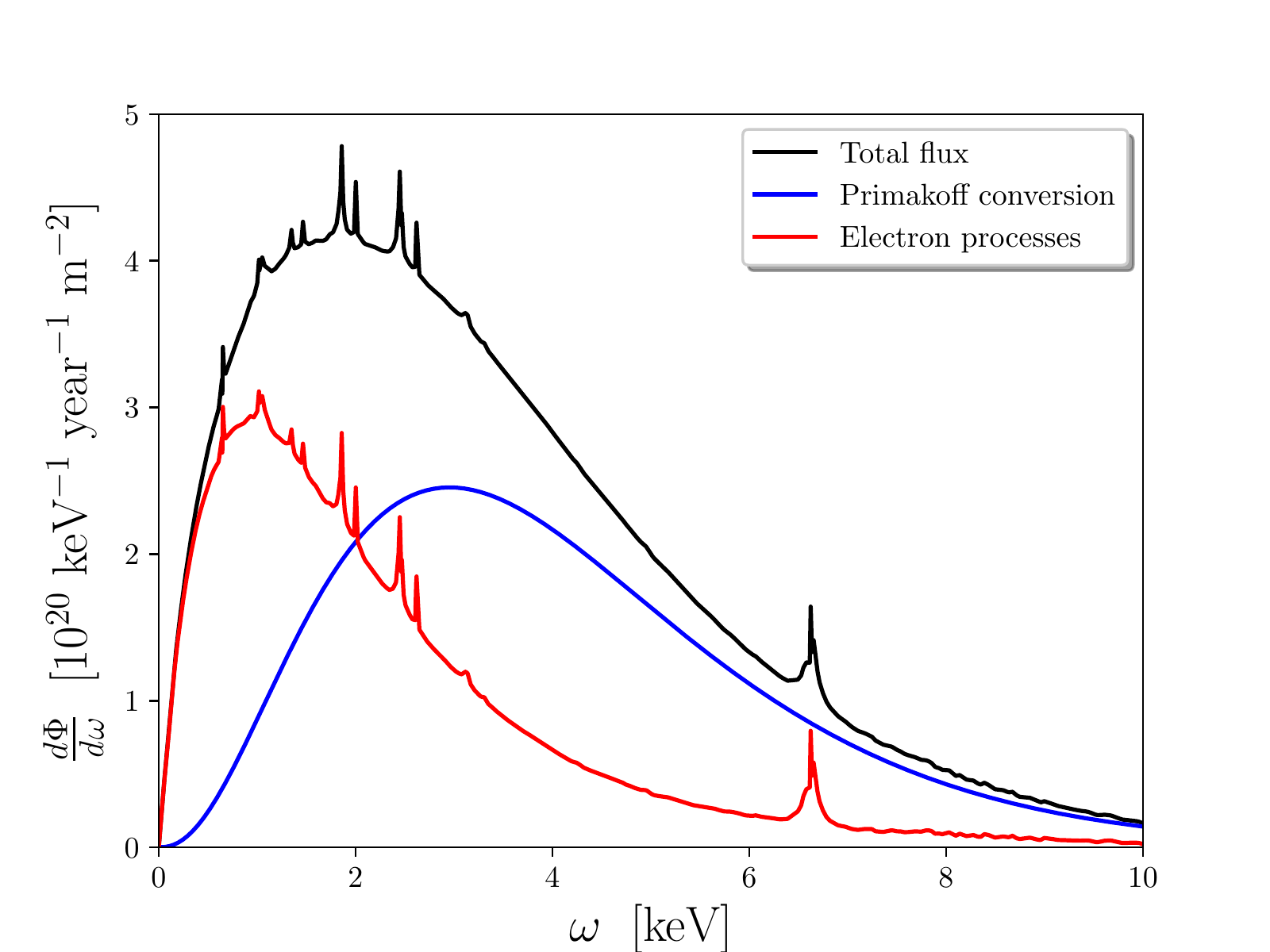}
\caption{The solar axion flux from Primakoff conversion due to the photon coupling $g_{a\gamma}$ only (blue), from the electron coupling only (red). The combined spectrum is shown as the black line. The curves correspond to ${g_{a\gamma}=10^{-11} \ \text{GeV}^{-1}}$ and ${g_{ae}=10^{-13}}$. For our purposes the most important observation is that the spectrum due to the electron coupling is softer than the Primakoff spectrum due to the photon coupling. The data for the spectra is taken from~\cite{Redondo:2013wwa}.}
\label{flux}
\end{figure}

The solar axion flux and its spectrum were calculated in \cite{Redondo:2013wwa}. We can divide the flux into two contributions: Primakoff production of axions from coupling to photons and effects requiring a coupling to electrons like axion-atomic transitions, axion-bremsstrahlung and the axion-Compton effect. The resulting fluxes are proportional to the squares of the two coupling constants $g_{a\gamma}^2$ and $g_{ae}^2$, respectively. Depending on the values of the coupling constants both contributions can play a significant role. 
The expected axion flux from these contributions is depicted in Fig.~\ref{flux}. For our aim of distinguishing the two contributions the most important feature is that the spectrum from couplings to electrons is much softer than the spectrum from Primakoff conversion only. The former peaks at approximately \SI{1.5}{\kilo \electronvolt} and the latter at \SI{3.5}{\kilo \electronvolt}. In addition, the bound-bound transitions of electrons cause distinct peaks at the transition energies. The detailed shape and width of these are under discussion, but it is clear that they will only be relevant if the axion couples to electrons. 

\bigskip
Let us now turn to the spectral features of the helioscope itself.
The solar axions enter the helioscope carrying the energy $\omega$.  
Starting from the Lagrangian given in Eq.~\eqref{lagrangian}, one can solve the classical equations of motion and compare the flux amplitudes to find the tree level result for the conversion probability~\cite{Sikivie:1983ip,Raffelt:1987im} of an axion to a photon in the conversion volume. The general result including a buffer gas is given by (cf.~\cite{Arik:2008mq}),
\begin{equation}
P_{a \rightarrow \gamma}= \left(\frac{g_{a\gamma}B}{2}\right)^2\frac{1}{q^2+\Gamma^2/4}\left[1 + e^{-\Gamma L}-2e^{-\Gamma L /2}\cos\left(qL\right)\right].
\label{conversion}
\end{equation} 
Here, $B$ is the transversal magnetic field strength, $L$ the length of the conversion volume, $q$ the transferred momentum and $\Gamma$ the inverse absorption length of photons in the gas. Only $q$ and $\Gamma$ depend on $\omega$. Defining $m_\gamma$ as the effective mass of a photon inside the gas, we can calculate $q=\frac{1}{2\omega}(m_a^2-m_\gamma^2) $. For a $^4$He buffer gas $m_\gamma$ can be approximated as \cite{Galan},
\begin{equation}
m_\gamma \approx \sqrt{0.02 \ \frac{p(\text{mbar})}{T(\text{K})}} \ \text{eV},
\label{gammamass}
\end{equation}
where $p$ and $T$ are the pressure and the temperature of the gas. $\Gamma$ can be deduced from the energy dependent mass attenuation coefficient $\frac{\Gamma}{\rho}$ by multiplying with the density of the gas $\rho$. An approximation of $\Gamma$ (based on a fit to the data~\cite{nist}) in the interesting energy range is given by,
\begin{equation}
\Gamma\approx 0.29\times\frac{p(\text{mbar})}{\omega(\text{keV})^{3.1}\ T(\text{K})}\text{m}^{-1}.
\end{equation} 
However, in our simulation, we take the full listed values from~\cite{nist}.
In the vacuum case $\Gamma \rightarrow 0$ and $q \rightarrow \frac{m_a^2}{2\omega}$. 

In addition to the conversion probability, we have to include the combined efficiency of the X-ray window at the end of the gas system, the X-ray optics and the detector itself. We call this factor $Q(\omega)$. For our purposes, this factor can be crucial in setting the energy threshold of the system. To achieve a low energy threshold, both the detector as well as the X-ray optics need to be suitably designed.

The total spectral flux of detected photons is,
\begin{equation}
\dv{\Phi_\gamma}{\omega}=Q(\omega)P_{a \rightarrow \gamma}(\omega)\dv{\Phi_a}{\omega}.
\end{equation}
  
\section{Simulating IAXO with energy resolution}\label{simulation}
Let us now describe our method for finding the parameter space in which separate measurements of the two coupling constants is possible. 
\subsection{Massless case}
\label{masslesssection}
In the limit of a massless axion and at vanishing pressure in the helioscope, both the transferred momentum $q$ and the inverse absorption length $\Gamma$ are set to zero. This strongly simplifies Eq.~\eqref{conversion} to,
\begin{equation}
P_{a \rightarrow \gamma}= \left(\frac{g_{a\gamma}BL}{2}\right)^2,
\end{equation}
which is now independent of $\omega$. We further assume the efficiency $Q$ to be $\frac{1}{2}$ for all energies up to the maximal energy of \SI{10}{\kilo \electronvolt}. With this broad simplification the only spectral dependence comes from the solar axion flux $\dv{\Phi_a}{\omega}$ but a more realistic description of the experimental setup could easily be included. This does not crucially affect the general strategy presented here.

We now assume for simplicity that the detector has $N$ energy bins in the range $[E_{\rm min},E_{\rm max}]$ with $\Delta E=E_{\rm max}-E_{\rm min}$. The expected number of counts in the $n$-th bin $\mu_n$ is given by,
\begin{equation}
\mu_n=QP_{a \rightarrow \gamma}\int_{\omega_{n-1}}^{\omega_{n}}\dv{\Phi_a}{\omega}\mathrm{d}\omega \times t\times A + \mu_{\text{b}},
\label{mu}
\end{equation} 
with
\begin{equation}
\omega_n=\left(E_{\rm min}+\frac{\Delta E}{N}n\right).
\end{equation} 
Here, $t$ is the total observation time and $A$ the effective cross-section of the helioscope. 
$\mu_{\text{b}}$ are the expected background counts, which we assume to be the same in every bin. 
For concreteness we take the background level as,
\begin{equation}
\mu_{b}=(1 \times 10^{-7} \frac{1}{\text{s keV cm}^2})\times A_\text{detect}\times t\times\frac{\Delta E}{N}\,.
\end{equation}
$A_\text{detect}$ is the detector area on which all the photons are focused.  According to~\cite{Armengaud:2014gea}, this is a realistic value for the background level. It will turn out that in the parameter space of interest to us, this corresponds to almost zero background (less than 1\% background events).

We take the numerical results of \cite{Redondo:2013wwa} and use Poissonian statistics to simulate a binned signal for IAXO. To recover the coupling parameters entering the simulation, we apply the same maximum likelihood method as in \cite{Barth:2013sma}. The number of counts $c_n$ in each energy bin are $N$ independent Poissonian variables, for which a likelihood function can be defined as,
\begin{equation}
\mathcal{L}= \prod_{n=1}^N\frac{e^{-\mu_n}\mu_n^{c_n}}{c_n!}.
\end{equation}
The $\mu_n$'s depend on the parameters entering the simulation, which we combine and call $\lambda$. The best fit value $\lambda ^*$ can be found by minimizing $\chi^2=-2 \log (\mathcal{L})$. A likelihood-ratio test was used to find an approximate 95\% certainty interval. The parameter $\lambda$ was accepted when the log-likelihood ratio,
\begin{equation}
\Lambda= -2 \log \left(\frac{\mathcal{L}(\lambda |c_1...c_n)}{\mathcal{L}(\lambda^* |c_1...c_n)}\right),
\end{equation} was smaller than the 95th percentile of the $\chi^2$-distribution with $k$ degrees of freedom. Here, $k$ is the number of free parameters. We ensured that a 95\% certainty was reached by comparing the accepted values with the true input parameters for a large number of different simulated signals.  

\begin{figure}
\centering
\makebox[\textwidth][c]{\includegraphics[width=1.1\linewidth]{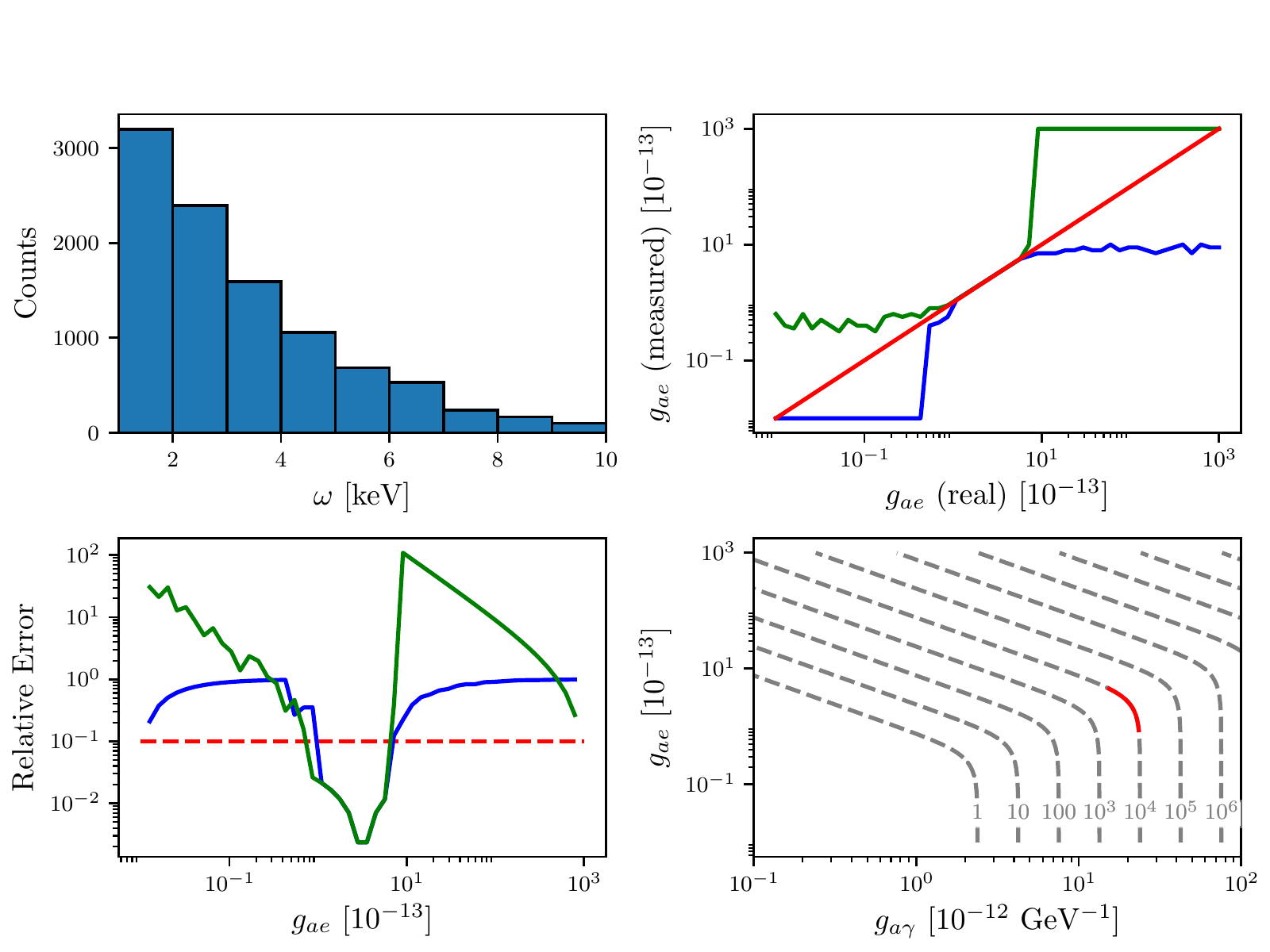}}
\centering
\caption{\textit{Top left panel:} An example of a simulated signal with an energy resolution of \SI{1}{\kilo \electronvolt} with $g_{ae}= 3.5 \times 10^{-13}$. \textit{Top right panel:} 95$\%$ likelihood intervals $g_{ae}$ (we have simulated 50 different values of $g_{ae}$ in the depicted range). Red is the true value entering the simulation, green the upper and blue the lower boundary of the deduced interval.  \textit{Bottom left panel:} Relative errors of upper (green) and lower boundary (blue). The limit at which the coupling is regarded as resolved is chosen as 10$\%$ and shown in red.  \textit{Bottom right panel:} The corresponding region in parameter space (red) is a line along the lines of constant number of events (dashed grey). }
\label{procedure}
\end{figure}

In a two-dimensional parameter space spanned by $g_{a\gamma}$ and $g_{ae}$, we immediately see that the direction of degeneracy is always parallel to the lines of constant number of overall counts (grey lines in the bottom right panel of Fig.~\ref{procedure}). Sensitivity curves of helioscopes (like the ones in \cite{Irastorza:2013dav,Barth:2013sma}) also follow these lines. In the Primakoff dominated case they are parallel to constant $g_{a\gamma}$ and in the electron coupling dominated case parallel to constant $g_{a\gamma}g_{ae}$. The most promising place to break this degeneracy is in the vicinity of comparable fluxes. In Fig.~\ref{procedure} this happens where the grey lines form a ``knee''.

Orthogonal to the lines of constant flux the accuracy of the measurement is very good and easy to understand from Poissonian statistics. The lowest number of total counts $\mu$ we are considering is a few hundred. In this case the relative error for $\mu \sim g^4$ is approximately 5\%. This means that the relative error for the couplings does not become much larger than 1\% in the relevant parameter space. Therefore, we restrict ourselves in the following to lines of constant expectation value and thereby reduce our parameter space to a one-dimensional one, which can be parametrised by $g_{ae}$, while $g_{a\gamma}(\mu,g_{ae})$ is always chosen such that the total expected number of counts remains constant.

Our procedure is illustrated in Fig.~\ref{procedure}. We start by simulating a IAXO-signal from Poissonian statistics in every bin for 50 values of $g_{ae}$ ranging from $10^{-15}$ to $10^{-10}$, but with a constant expectation value for the number of events. An example of such a signal is depicted in the top left panel of Fig.~\ref{procedure}. For every value the 95$\%$ certainty interval for $g_{ae}$ is calculated using the likelihood-ratio test described above (top right panel). The relative errors are calculated for the maximal and the minimal accepted values (bottom left panel). If both are smaller than 10$\%$, we regard the two couplings as individually resolved. The resulting line can be plotted in parameter space (bottom right panel). By averaging over 20 repetitions of this procedure, statistical fluctuations are reduced and we arrive at the expected sensitivity. This was done for a large number of different expectation values in order to find the area in parameter space in which $g_{ae}$ can be resolved with at least 10$\%$ accuracy. Because the number of counts and thereby $\mu$ can be resolved well, we can deduce $g_{a \gamma }$. 
The relevant parameters entering the simulation are shown in Tab.~\ref{parameters}. In agreement with the preliminary IAXO design~\cite{Armengaud:2014gea}, the figure of merit was chosen 300 times larger than the one of the CAST experiment. IAXO is designed for 12 hours of sun tracking a day. This should make it possible to achieve the 100 days of tracking within one year. We take this as our benchmark value for the measurement time, $t$.
\begin{table}
\centering
\renewcommand{\arraystretch}{1.2}
\begin{tabular}{|l|c|}
\hline
Parameter & Value \\
\hline
Magnetic field strength $B$ & \SI{2.8}{\tesla} \\
Length of conversion volume $L$ & \SI{20}{\metre} \\
Cross-section of conversion volume $A$ & \SI{2}{\square \metre} \\
Figure of merit ($B^2L^2A$)&\SI{6272}{\square \tesla \metre \tothe{4}} ($\sim 300 \ \times$ CAST)\\
Total tracking time $t$ & \SI{100}{days} \\
Bandwidth & \SIrange[range-units = brackets]{1}{10}{\kilo \electronvolt} \\
Energy resolution $\Delta\nu$ & \SI{1}{\kilo \electronvolt} \\
Inverse absorption length $\Gamma$ & 0 (vacuum)\\
Efficiency of telescope $Q$ & 0.5 \\
Background level & \SI{e-7}{\per \kilo \electronvolt \per \second \per \square \cm}\\
Detector area $A_\text{detect}$ & \SI{1}{\square \cm}\\
\hline
\end{tabular}
\caption{Parameters entering our simulation for the massless case. We will later discuss some possible optimisation especially for the bandwidth and the energy resolution. These values are all based on the preliminary IAXO design \cite{Armengaud:2014gea}.  }
\label{parameters}
\end{table}

\begin{figure}
	\centering
	\includegraphics[width=.6\linewidth]{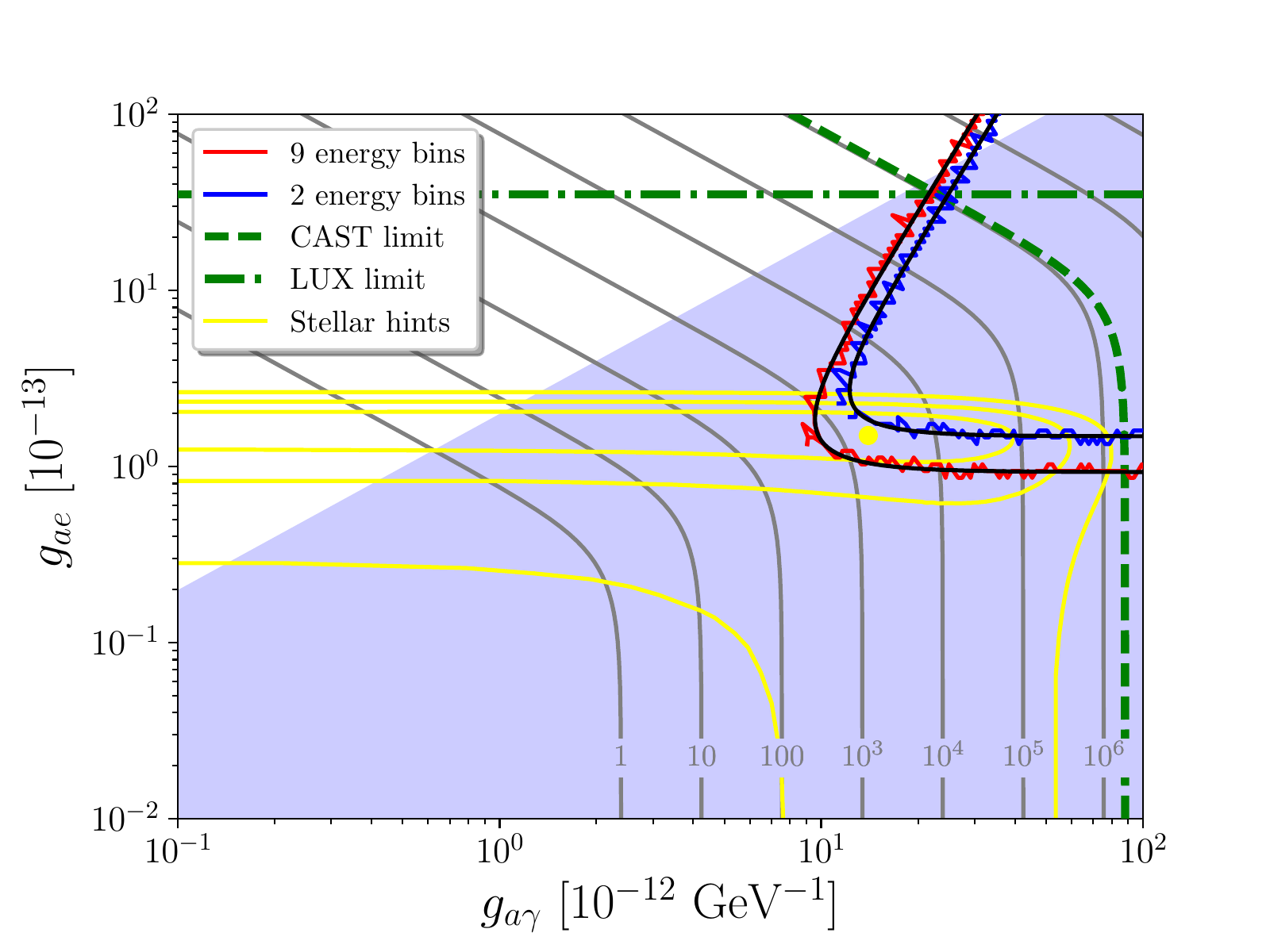}
	\caption{Border of parameter space in which $g_{a\gamma}$ and $g_{ae}$ can be measured individually with \SI{1}{\kilo \electronvolt} and \SI{4.5}{\kilo \electronvolt} energy resolution (red and blue lines). The larger area (red line) was achieved with better resolution. The numerical results were fitted with an appropriate function (black lines). Parameter space covered by the DFSZ I and II models is shaded in blue. Above this region, perturbativity is violated.~\cite{Irastorza:2018dyq,Chen:2013kt,Giannotti:2017hny} Each point in coupling space has a corresponding model-dependent mass which we ignore here. Hints from stellar cooling~\cite{Giannotti:2017hny} are shown in yellow. The best fit value as well as 1, 2 and 3~$\sigma$ likelihood intervals are depicted. Larger couplings than the 3~$\sigma$ contour can be considered disfavoured by stellar cooling. Experimental limits of the CAST \cite{Barth:2013sma} and LUX~\cite{Akerib:2017uem} (similar but slightly weaker limits were produced by XENON100~\cite{Aprile:2014eoa} and PandaX~\cite{Fu:2017lfc}) experiments are plotted as well. The grey lines correspond to lines of constant number of events starting from one event at the bottom left and going up in powers of ten.}
	\label{massless}
\end{figure}

The results for \SI{1}{\kilo \electronvolt} resolution (number of bins $N=9$) and for a distinction between high and low energies ($N=2$) can be seen in Fig.~\ref{massless}. 
For our chosen helioscope setup and running time a better resolution than \SI{1}{\kilo \electronvolt} does not improve this result dramatically. 
The main reason for this is the overall quite small number of events for the setup we investigate. Indeed, the peaks from bound-bound state transitions do not have sufficient additional statistical weight. For most of the relevant coupling strengths the number of events contributed by the peaks is low and they are either invisible (large number of events) or indistinguishable from statistical fluctuations of the continuous part of the spectrum (low number of events).
However, we will see in the following that a lower threshold can be beneficial for the coupling resolution. Moreover, a better energy resolution allows to break degeneracies in the massive case.
  
The asymptotic behaviour of the boundaries can also be understood analytically. For large electron couplings the contribution from Primakoff production approximately becomes relevant when the corresponding number of events is comparable to the standard deviation of the dominant electron coupling contribution,
\begin{equation}
g_{a\gamma}^4\sim \sqrt{g_{a\gamma}^2g_{ae}^2} \Rightarrow g_{ae} \sim g_{a\gamma}^3.
\end{equation}
This defines the border between the region dominated by the flux from coupling to electrons and the one with both contributing significantly.
Similarly, for the border to the Primakoff dominated region we have,
\begin{equation}
g_{a\gamma}^2g_{ae}^2\sim \sqrt{g_{a\gamma}^4} \Rightarrow g_{ae} \sim \text{const}.
\end{equation}
This is exactly the behaviour we observe, and our fit in Fig.~\ref{massless} converges to the two gradients 3 and 0 within 1 $\sigma$. The dimensionful proportionality constants depend on the statistical details, such as the certainty limit.  

\begin{figure}
	\centering
	\includegraphics[width=.6\linewidth]{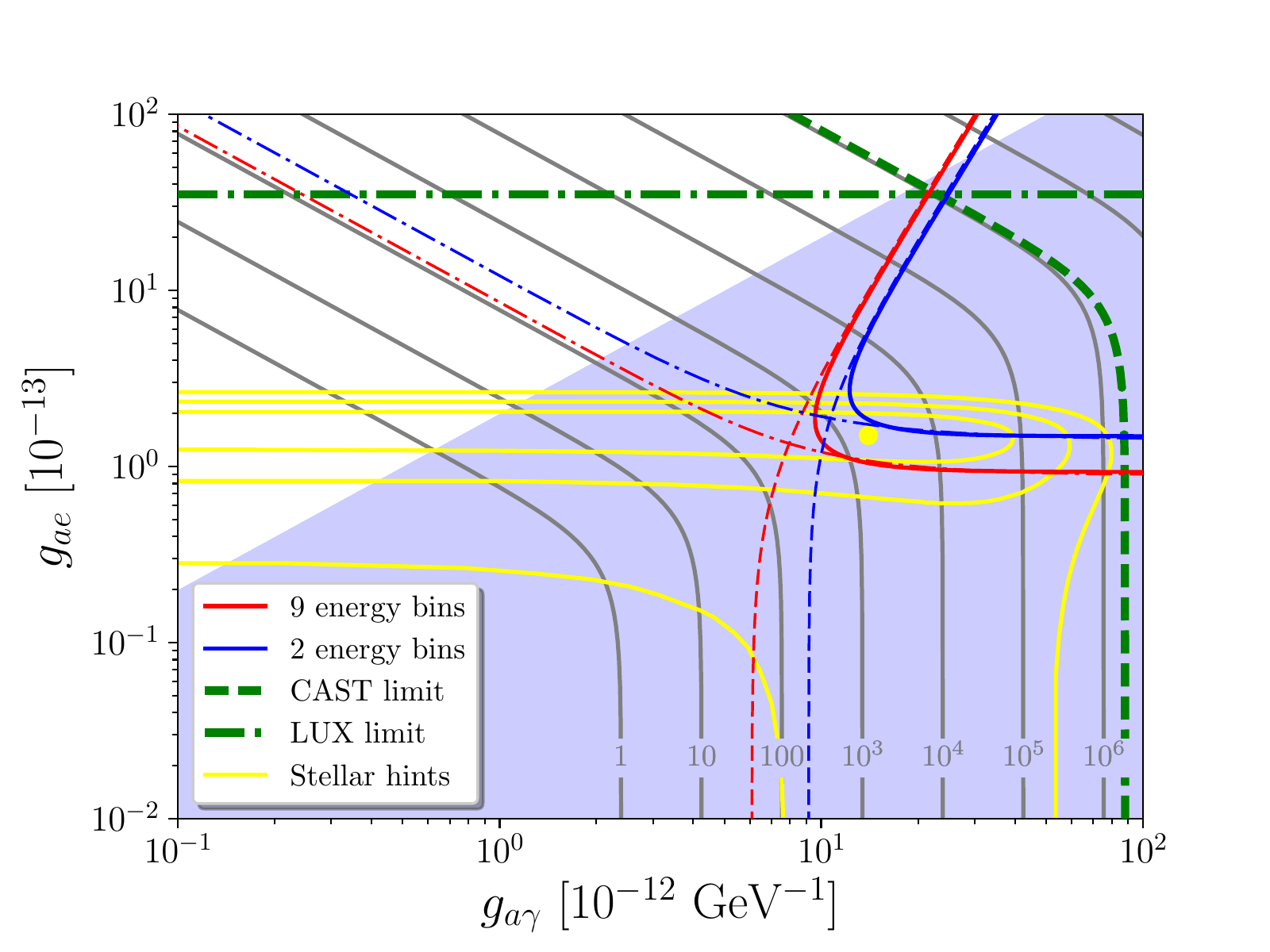}
	\caption{Borders of parameter space in which $g_{a\gamma}$ (dashed lines) or $g_{a\gamma}g_{ae}$ (dash-dotted lines) can be measured  with 10\% accuracy in setups with \SI{1}{\kilo \electronvolt} and \SI{4.5}{\kilo \electronvolt} energy resolution. The solid lines are the same as in Fig.~\ref{massless}. If it is not possible to measure both, there is an upper bound for $g_{ae}$ or $\frac{g_{a\gamma}}{g_{ae}}$. For simplicity only the fits of the numerical data are shown. Other elements of the plot are equivalent to the ones in Fig.~\ref{massless}.}
	\label{oneparameter}
\end{figure}
In Fig.~\ref{massless}, we include the best region for the stellar cooling anomalies found in~\cite{Giannotti:2017hny}. It is interesting that a large part of this region allows to resolve the electron and photon couplings.
Moreover, it covers a sizeable part of the parameter space accessible to DFSZ models.

Even in cases where it is not possible to distinguish the two couplings individually, additional information can be gained from energy resolved detection. In Fig.~\ref{oneparameter}, we show the regions where either $g_{a\gamma}$ or $g_{a\gamma}g_{ae}$ can be measured with 10\% accuracy. Again, we compare two setups with different energy resolutions. As expected, the overlap roughly\footnote{The small difference at the tip of the overlap region arises from the fact that measuring $g_{a\gamma}$ and $g_{ae}$ to a given precision is not exactly equivalent to measuring $g_{a\gamma}$ and $g_{a\gamma}g_{ae}$ to the same precision. This can be seen from the propagation of errors.} corresponds to the previously found region plotted in Fig.~\ref{massless}. If only $g_{a\gamma}$ can be measured (Primakoff dominated case), there will be an upper bound for $g_{ae}$. In the case with axion-electron domination, the product of the two couplings can be measured and an upper bound on $\frac{g_{a\gamma}}{g_{ae}}$ can be extracted.

\subsection{Optimising the setup by lowering the detection threshold}
Let us consider a possible optimisation of the setup that would improve the sensitivity in distinguishing between the two couplings.

\begin{figure}
\centering
\includegraphics[width=.6\linewidth]{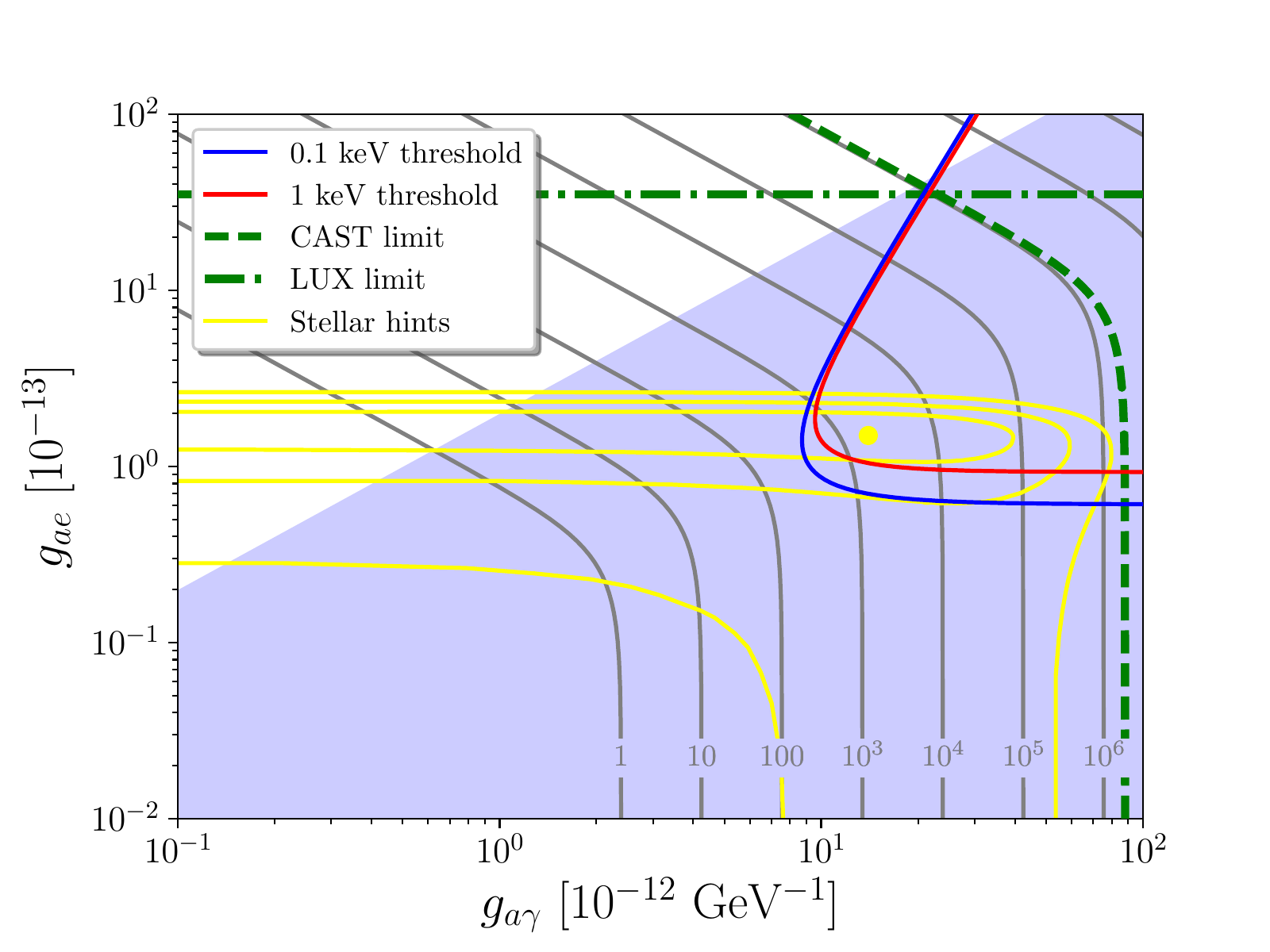}
\caption{Comparison of the result for higher (red line) and lower (blue line) detection threshold for a massless axion. The energy resolution is \SI{1}{\kilo \electronvolt} in both cases. We use the same method as in the massless case. As expected, the lower threshold enables detection of an even smaller number of photons from electron coupling processes because these contribute mostly in the low energy bins. For simplicity only (smoothed out) fitting curves are depicted. Everything else as in Fig.~\ref{massless}.}
\label{lowthresh}
\end{figure}

From the spectrum in Fig.~\ref{flux}, it is clear that the optics in the preliminary IAXO design (\SIrange[range-units = single]{1}{10}{\kilo \electronvolt}) is not optimised for the axion-electron flux contribution to the photon flux since a significant fraction of the axion-electron flux is below \SI{1}{\kilo \electronvolt}. Instead, it would be preferable to have the ability to detect lower energy photons. While this will only slightly enlarge the region in which the axion can be discovered, it significantly increases the parameter space in which separate measurements of the couplings are possible. To quantify this, the sensitivity for detection is only improved by a factor of about 1.15 in the case of an electron coupling dominated flux. In contrast, the region where we can measure the couplings is enlarged towards weaker electron coupling by a factor of $\sim$1.5.
The reason for this is that even a few counts in the lowest energy bins can indicate non-vanishing coupling to electrons. In Fig.~\ref{lowthresh}, our previous result for a massless axion is compared to one with a detection threshold of just \SI{0.1}{\kilo \electronvolt}. 
This significant improvement provides motivation to try to optimise IAXO's optics and detectors towards lower energies.

\subsection{Using the buffer gas for energy resolution}
On the other hand, even without any energy resolution, it is possible to gain some information about the different couplings. It was already pointed out by the CAST collaboration~\cite{Arik:2008mq} that by detuning the pressure of the buffer gas from the resonant value, $q$ gets a non-zero value and $P_{a \rightarrow \gamma}$ becomes energy dependent. More specifically, for small detuning from resonance the softer photons are strongly suppressed.  This allows to achieve an effective energy resolution by using a buffer gas.

We can use this to do a simulation with only one energy bin in the detector, but with one observation on resonance and one slightly off resonance ($m_\gamma=$\SI{15}{\milli \electronvolt}). The new likelihood function is the product of the two runs,
\begin{equation}
\mathcal{L}=\frac{e^{-\mu}\mu^{c}}{c!} \times \frac{e^{-\tilde{\mu}}\tilde{\mu}^{\tilde{c}}}{\tilde{c}!},
\end{equation}
where $\mu$ and $\tilde{\mu}$ are the two expectation values on and off resonance, while $c$ and $\tilde{c}$ are the actual numbers of events. We do the same steps as before to derive the parameter space in which separate measurements of the two relevant couplings is possible. For illustration we focus on the massless case.

\begin{figure}
\centering
\includegraphics[width=.6\linewidth]{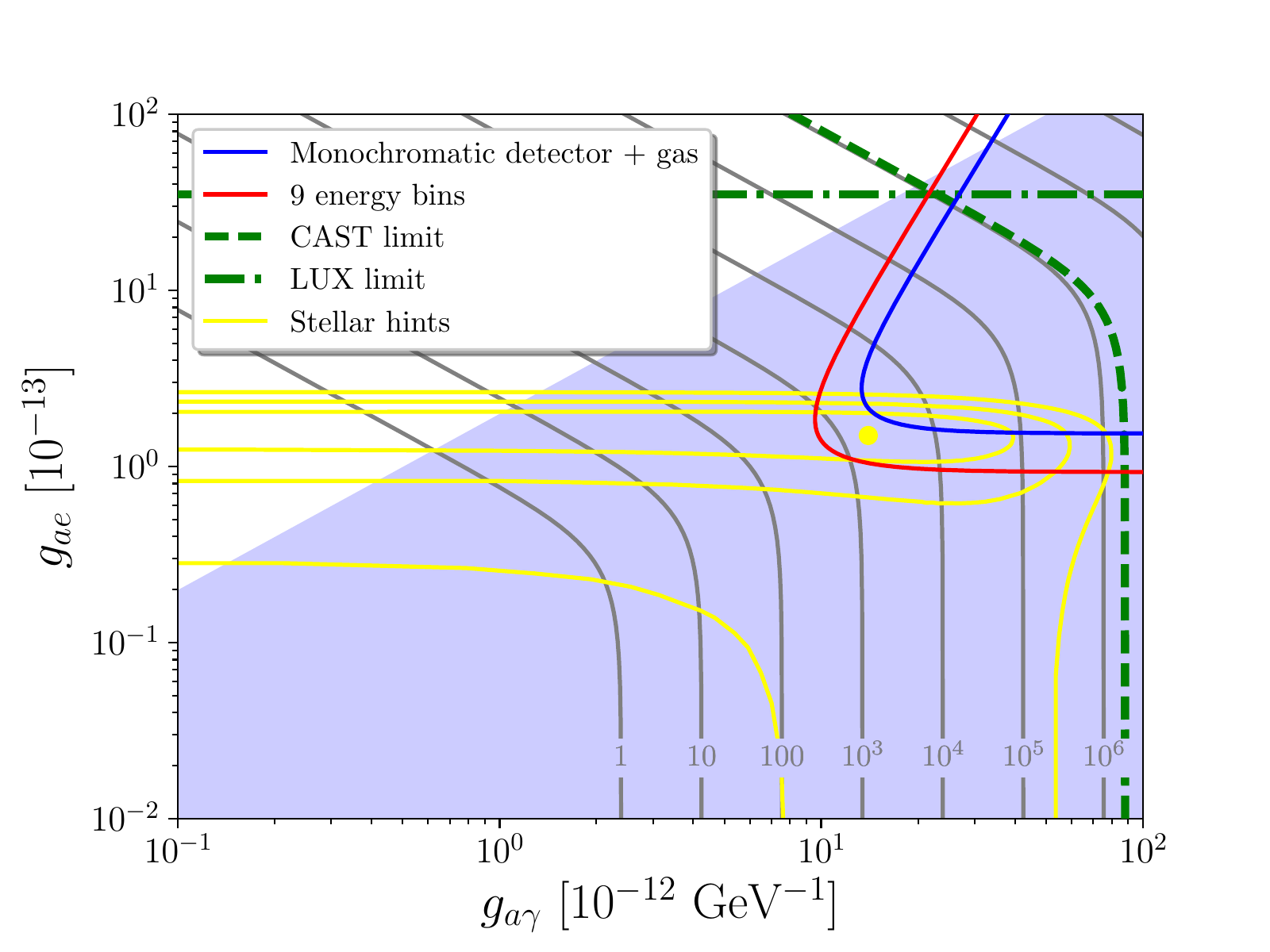}
\caption{Comparison of energy resolved detection with \SI{1}{\kilo \electronvolt} energy resolution (red line) and monochromatic detection at different pressure settings of the buffer gas (blue line). The latter allows individual measurements of $g_{a\gamma}$ and $g_{ae}$ in a significantly smaller region of parameter space. For simplicity only the fits are depicted. All other features are as in Fig.~\ref{massless}.}
\label{gasres}
\end{figure}

While this method can provide some information, it cannot compete with good energy resolution as can be seen from Fig.~\ref{gasres}.
In particular, we lose a significant part of the best fit parameter region for the stellar cooling anomalies.

\subsection{Massive case}

Allowing for a non-vanishing mass adds an additional parameter to be measured. It also adds a few complications to the previous scenario. For higher masses than approximately \SI{5}{\milli \electronvolt} the conversion probability $P_{a \rightarrow \gamma}$ given in Eq.~\eqref{conversion} starts to oscillate and drop rapidly. The reason is the decoherence between the photon and the axion wave. This effect is most pronounced at low axion energy because there the difference in the dispersion relation due to the mass is bigger. This effect in principle enables us to measure the mass \cite{Arik:2008mq,Irastorza:2018dyq}. 

\begin{figure}[!t]
\centering
\includegraphics[width=1.\linewidth]{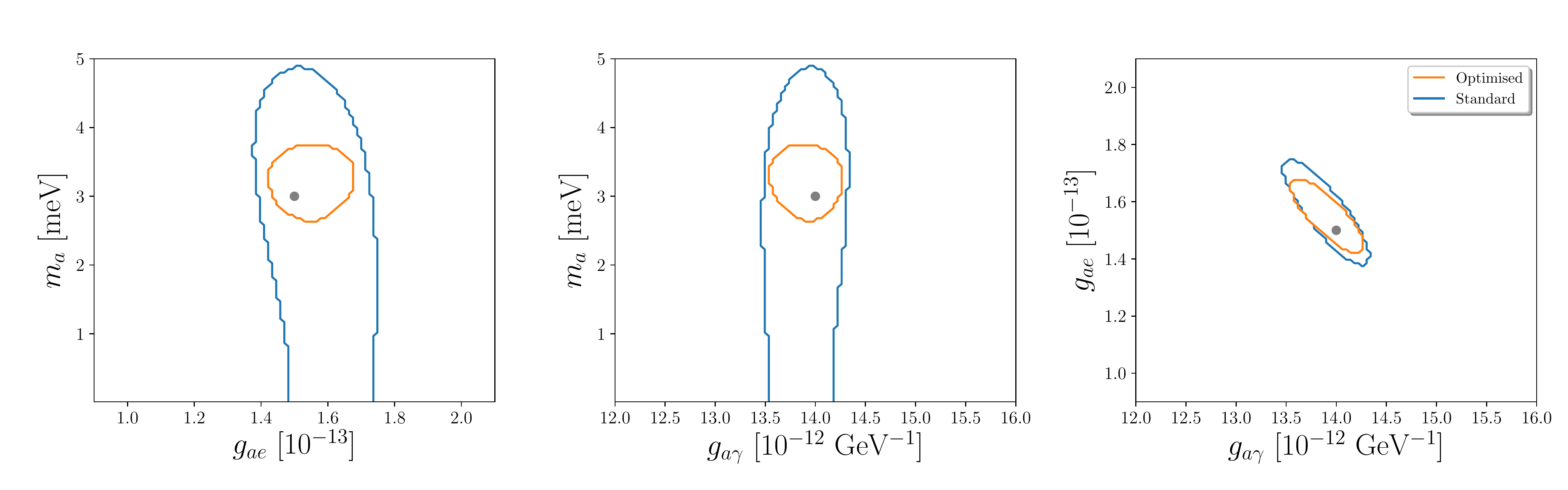}
\caption{A single simulation of case 1 in Tab.~\ref{best_fit_case}. $m_a$ was chosen as \SI{3}{\milli \electronvolt}. The three panels show the 95\% certainty interval projected onto three different planes in parameter space. For this, two long ($t=100 \ \text{days}$) observations -~one without a buffer gas and one with a gas tuned slightly off resonance~- were simulated. We can clearly see that the optimised setup (threshold \SI{0.1}{\kilo \electronvolt} and energy resolution of \SI{0.1}{\kilo \electronvolt}) is able to resolve $m_a$, while the standard setup (threshold \SI{1}{\kilo \electronvolt} and energy resolution \SI{1}{\kilo \electronvolt}) is not. This is mainly due to the lower detection threshold. }
\label{case1}
\end{figure}

\begin{figure}[!t]
\centering
\includegraphics[width=1.\linewidth]{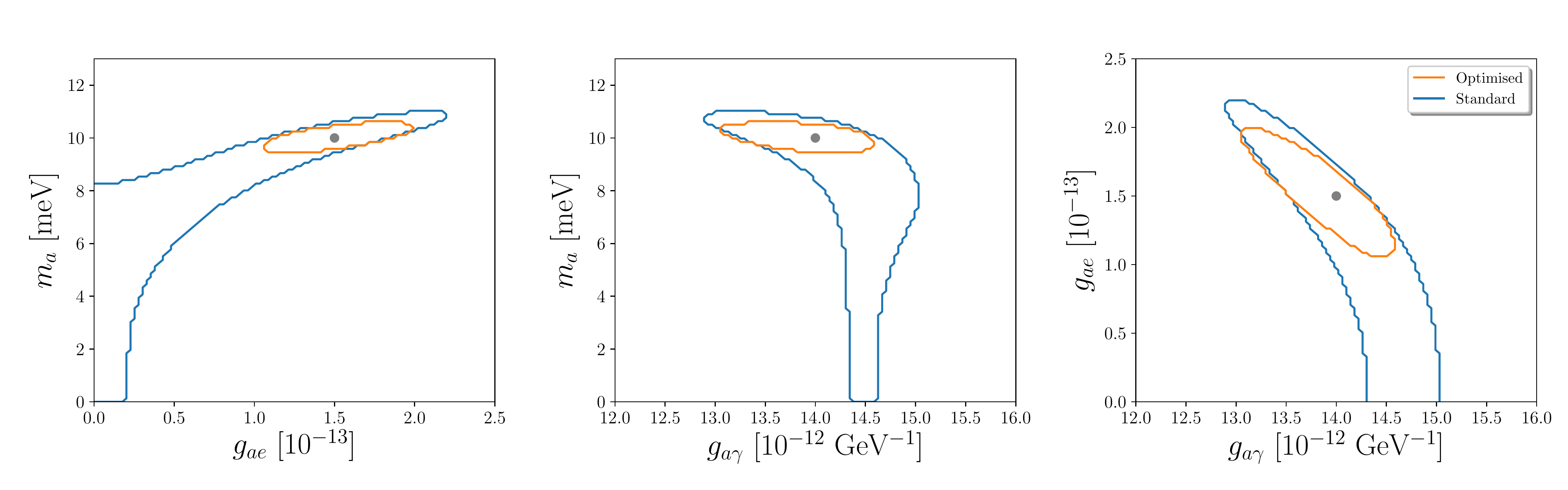}
\caption{A single simulation of case 2 in Tab.~\ref{best_fit_case}. $m_a$ was chosen as \SI{10}{\milli \electronvolt}. The three panels show the 95\% certainty interval projected onto three different planes in parameter space. Only one long ($t=100 \ \text{days}$) measurement without a buffer gas was simulated. Again, only the optimised setup can resolve all three different parameters. Here, both the low detection threshold and the good energy resolution are required.}
\label{case2}
\end{figure}

\begin{figure}[!t]
\centering
\includegraphics[width=1.\linewidth]{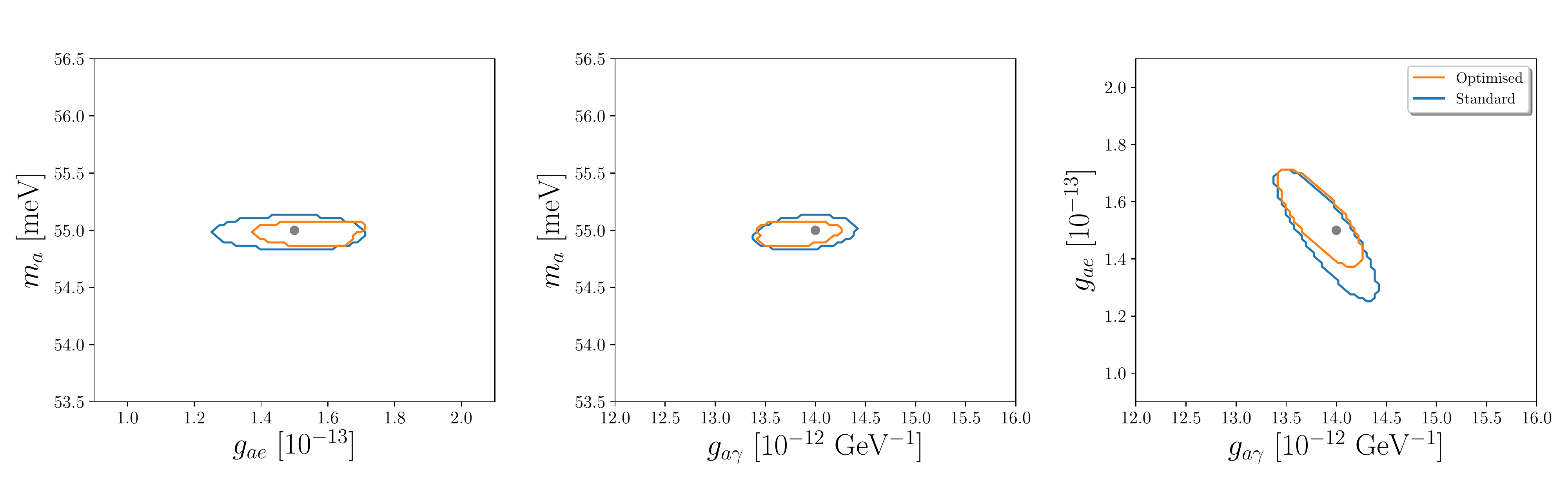}
\caption{A single simulation of case 3 in Tab.~\ref{best_fit_case}. $m_a$ was chosen as \SI{55}{\milli \electronvolt} which is the mass of a QCD axion in the DFSZ~II model with the couplings as in Eq.~\eqref{best_fit}. The three panels show the 95\% certainty interval projected onto three different planes in parameter space. We simulated a scanning of $m_\gamma$ in steps of \SI{2}{\milli \electronvolt} followed by two long ($t=100 \ \text{days}$) measurements at the two settings with the highest number of events during the scan. Here, the advantage of the optimised setup is relatively small. In both cases, we are able to resolve all three parameters individually.}
\label{case3}
\end{figure}

\begin{figure}[!t]
\centering
\includegraphics[width=1.\linewidth]{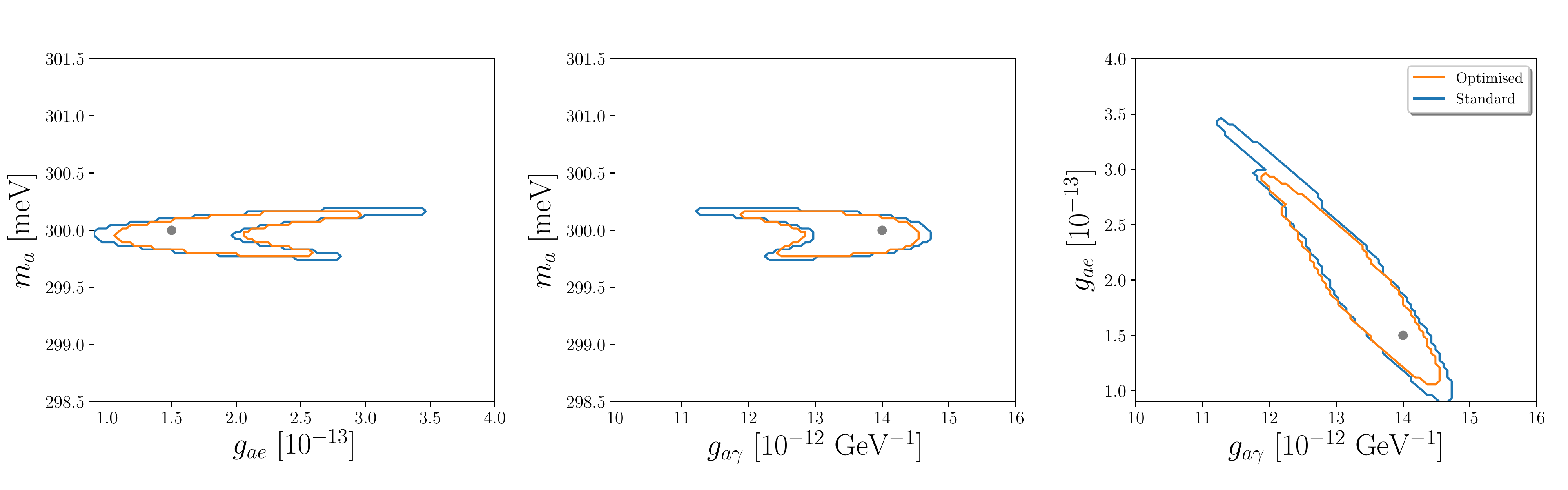}
\caption{A single simulation of case 4 in Tab.~\ref{best_fit_case}. $m_a$ was chosen as \SI{300}{\milli \electronvolt}. The three panels show the 95\% certainty interval projected onto three different planes in parameter space. We simulated a scanning of $m_\gamma$ in steps of \SI{2}{\milli \electronvolt} followed by two long ($t=100 \ \text{days}$) measurements at the two settings with the highest number of events during the scan. Because of the high mass a larger pressure is needed which leads to a lower number of events due to absorption. We can see degeneracy tails in the mass vs.\ coupling directions. These arise because, at a given gas pressure, the effect of a mass mismatch to the gas can to some degree be compensated by an adjustment of the couplings. Note the different axis scales in the third panel compared to Fig.~\ref{case3}.}
\label{case4}
\end{figure}

There are several ways to do this, all suited to a different part of the three-dimensional parameter space $(g_{a\gamma},g_{ae},m_a)$. To illustrate this, we look at the example of the best fit couplings given in Eq.~\eqref{best_fit} and discuss what can be achieved depending on the true value of $m_a$. Tab.~\ref{best_fit_case} summarises the different scenarios. In each case, we do a three-dimensional likelihood ratio test similar to the one in Sect.~\ref{masslesssection}, but only for a single simulated observation campaign. The plots in Figs.~\ref{case1}~to~\ref{case4} are therefore only examples of a possible result in each case. We always compare the sensitivity of the standard setup (\SI{1}{\kilo \electronvolt} threshold and \SI{1}{\kilo \electronvolt} resolution) to an optimised one (\SI{0.1}{\kilo \electronvolt} resolution and \SI{0.1}{\kilo \electronvolt} threshold). \\

\begin{table}
\centering
\renewcommand{\arraystretch}{1.3}
\makebox[\textwidth][c]{
\begin{tabular}{|l|l|c|c|c|c|}
	\hline
	$\#$&$m_a$ &Detection& $m_a$ resolved& $(g_{a\gamma},g_{ae})$ resolved & Method \\
	\hline
	0&$\lesssim$ \SI{2}{\milli \electronvolt} & \cmark &\xmark & \cmark & vacuum only\\
	\hline
	1& $\sim$\SIrange[range-units = brackets]{2}{5}{\milli \electronvolt} & \cmark & \cmark & \cmark & on/off resonance \\
	\hline
	2&$\sim$\SIrange[range-units = brackets]{5}{20}{\milli \electronvolt} & \cmark & \cmark & \cmark & vacuum only  \\
	\hline
	3&$\sim$\SIrange[range-units = brackets]{20}{200}{\milli \electronvolt} & \cmark & \cmark & \cmark & scanning $m_\gamma$  \\
	\hline
	4&$\sim$\SIrange[range-units = brackets]{0.2}{1}{ \electronvolt} & \cmark & \cmark & \xmark & scanning $m_\gamma$ \\
	\hline
	5&$\gtrsim$\SI{1}{\electronvolt}& \xmark & \xmark & \xmark & -\\
	\hline 
\end{tabular}}
\caption{Summary of the axion parameters which IAXO can detect in the case of the best fit couplings (Eq.~\eqref{best_fit}). The mass ranges should be seen as estimates because they depend on the details of the actual setup. Here, we have chosen the optimised setup with a detection threshold \SI{0.1}{\kilo \electronvolt} and energy resolution of \SI{0.1}{\kilo \electronvolt}. In reality, the mass ranges in which each method is applicable overlap, but here we only show what we think is the best suited method for every mass.}	

\label{best_fit_case}
\end{table}
For very small masses ($\lesssim$\SI{2}{\milli \electronvolt}) there simply is no decoherence effect which could be observed. In this case, the axion is detected in the vacuum setup and, by applying the method described in Sect.~\ref{masslesssection}, it is possible to resolve the couplings. 

If the mass is just slightly higher ($\sim$\SIrange[range-units = brackets]{2}{5}{\milli \electronvolt}), there will still be no appreciable decoherence in the vacuum. But combining the vacuum measurement with another one with a buffer gas slightly off resonance ($m_\gamma \sim$\SI{8}{\milli \electronvolt}), enables us to find the mass anyway. This is because $P_{a \rightarrow \gamma} $ is sensitive to $q=\frac{1}{2\omega}(m_a^2-m_\gamma^2) $. So a non-vanishing mass reduces the decoherence effect in the measurement off resonance. Fig.~\ref{case1} shows a simulated result of this procedure. A low energy threshold is crucial for this to work. 

At approximately \SI{5}{\milli \electronvolt} (depending on the energy threshold), the decoherence starts to be strong enough to be observed in the vacuum. An energy resolution of roughly \SI{0.1}{\kilo \electronvolt} would make it possible to distinguish between the effects of a small electron coupling and a non-vanishing mass. Both cause a lack of low energy photons, but the latter generates a sharper cut-off. Additionally, the decoherence from a non-vanishing $m_a$ causes oscillations in the spectrum for large enough masses~\cite{Arik:2008mq}, which can only be observed with sufficient energy resolution. This is why in this case both a low energy threshold and optimised energy resolution of \SI{0.1}{\kilo \electronvolt} are required to find the mass (see Fig.~\ref{case2}). 

For a mass larger than $\sim$\SI{20}{\milli \electronvolt} the decoherence is too strong to distinguish all parameters of the axion in the vacuum setup. Instead, one has to scan over different settings of $m_\gamma$. We simulated this as well. Scanning over $m_\gamma$ in steps of \SI{2}{\milli \electronvolt} with 5 days observation time on each setting should make it possible to find an approximation of the mass. Longer observations at the two pressure settings with the strongest signal enable us to find all three parameters (see Fig.~\ref{case3}). 

\begin{figure}[!t]
\centering
\includegraphics[width=.6\linewidth]{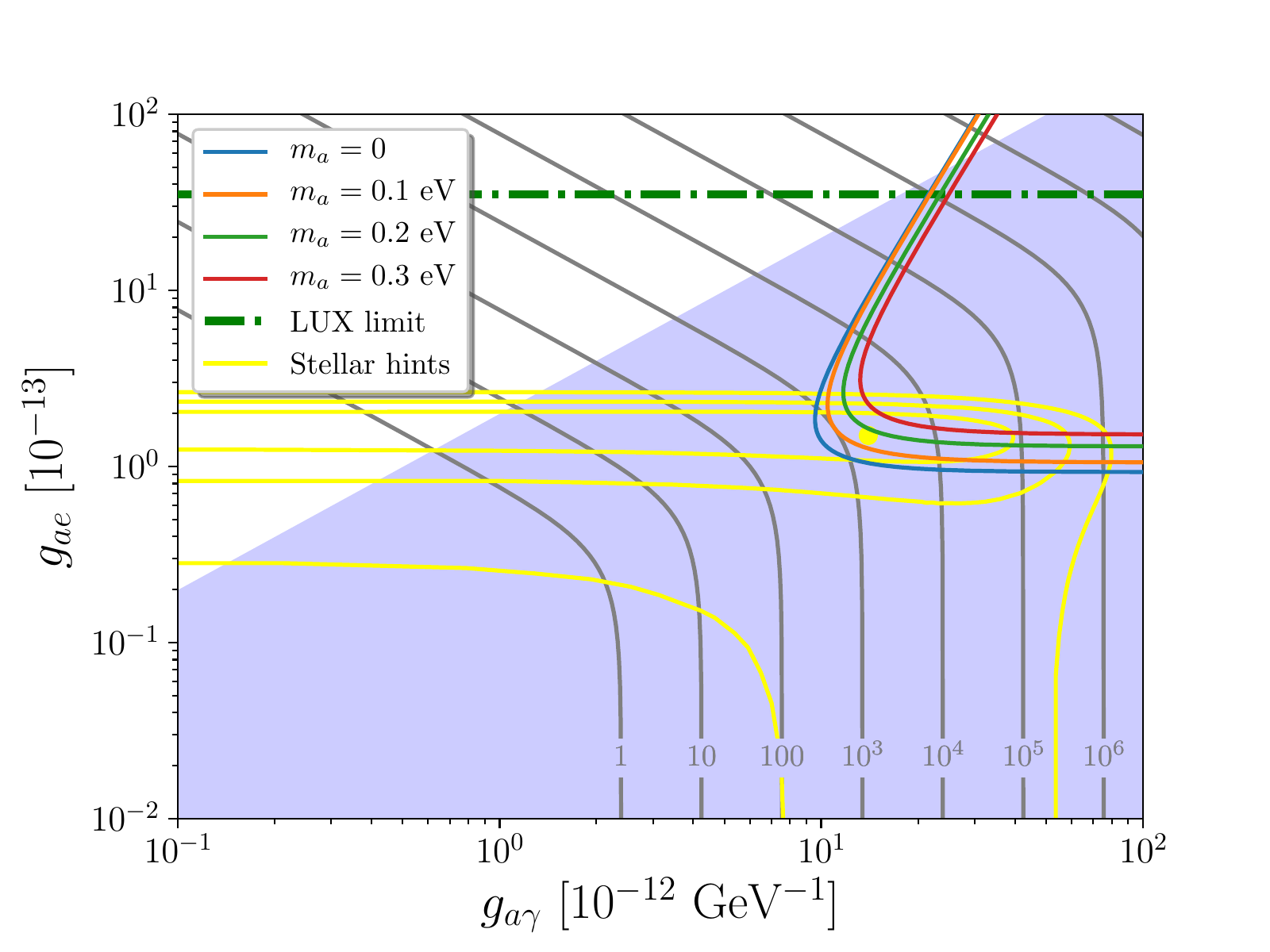}
\caption{Same as Fig.~\ref{massless} but for four different values of $m_a$ between 0 and \SI{0.3}{ \electronvolt}. The expected sensitivity decreases for higher masses (higher pressure and hence more absorption). Again, only the fits of the numerical results are depicted. At approximately \SI{0.2}{ \electronvolt} the best fit value exits the area with resolved couplings. The CAST limit does not appear anymore because its sensitivity is much lower for larger masses \cite{Arik:2008mq}.}
\label{const_mass}
\end{figure}

\begin{figure}[!t]
\centering
\includegraphics[width=.6\linewidth]{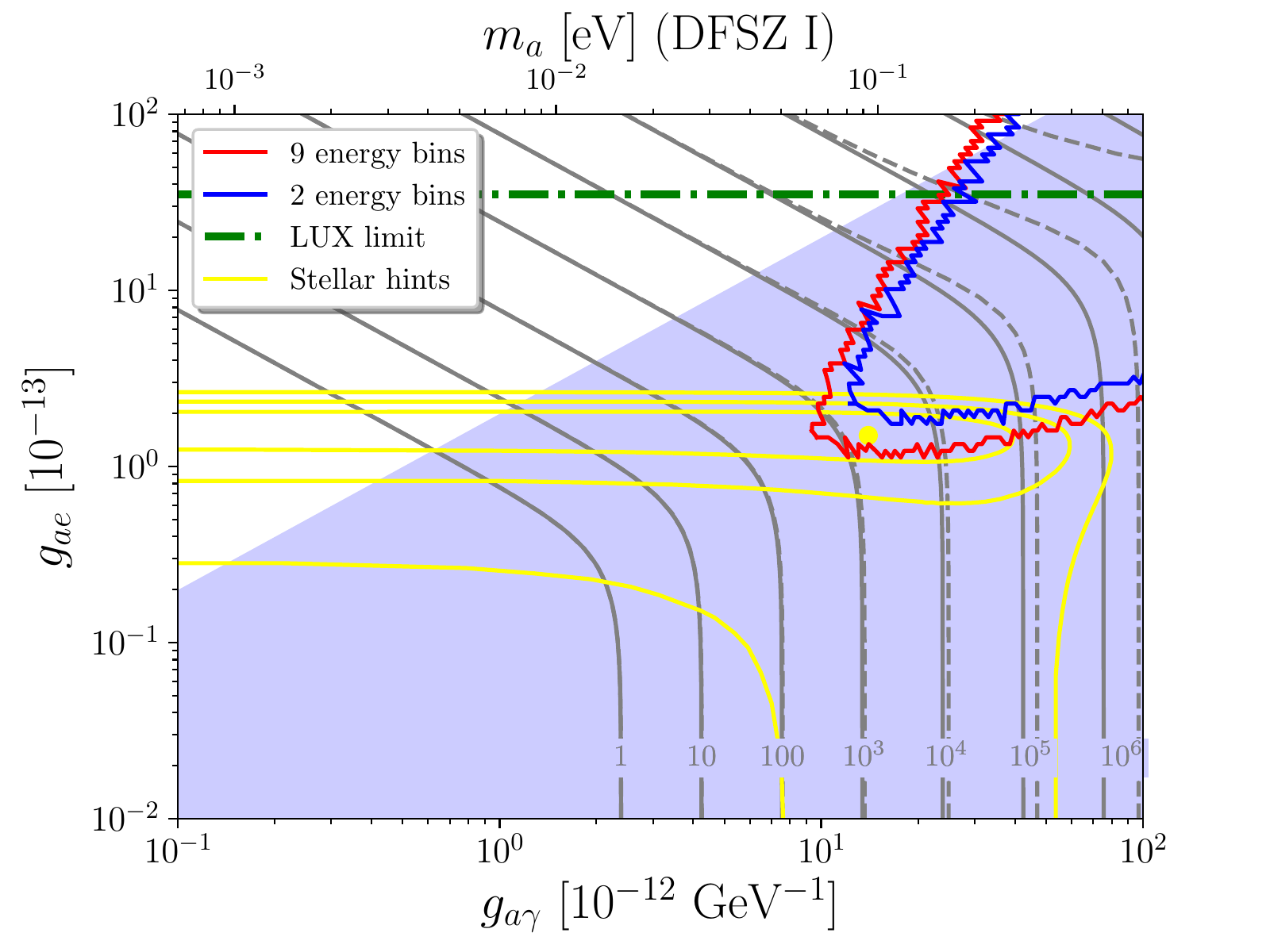}
\caption{Same as Fig.~\ref{massless} but with a mass corresponding to the DFSZ~I model. The new lines of constant number of events are shown in dashed grey and only deviate from the former solid grey ones for higher masses. The expected sensitivity (again for $N=2$ and $N=9$) decreases for these values (higher pressure and hence more absorption). But only very little parameter space compatible with stellar cooling is lost in comparison to Fig.~\ref{massless}. Because the proportionality between $g_{a\gamma}$ and $m_a$ spoils the asymptotic behaviour we did not fit the numerical results in this case. Again, the CAST limit does not appear in this plot because of too high masses.}
\label{massive}
\end{figure}

Even with a perfect pressure setting, the signal strength is reduced because of absorption in the gas. Therefore, higher masses always come with a smaller number of events and at some point the signal will not be strong enough to distinguish the two couplings. This can be seen by looking at a single simulation as in the other cases (see Fig.~\ref{case4}).\footnote{Note that we can still see that both couplings must be non-vanishing. While this is trivial for the photon coupling that is required for detection, this still is valuable extra information on the electron coupling.} Alternatively, Fig.~\ref{const_mass} illustrates that, for our benchmark couplings, we lose the ability to distinguish the couplings at roughly \SI{200}{\milli \electronvolt}. Again, this depends on the energy threshold. Fig.~\ref{massive} shows that, if there is a proportionality between $m_a$ and $g_{a\gamma}$ like in QCD axion models, we are still able to find the different couplings for a large part of the parameter space hinted at by stellar cooling. 

Finally, at approximately \SI{1}{ \electronvolt} the gas starts to condense and no further scanning is possible, making a detection impossible. The exact value at which this happens depends on the buffer gas. The CAST experiment used $^3$He to go beyond \SI{1}{ \electronvolt} \cite{Arik:2008mq}. 


\newpage
\section{Conclusions}\label{summary}
Using energy resolved detectors in axion helioscopes such as IAXO~\cite{Armengaud:2014gea} opens up the opportunity to gain additional information from a putative signal.
In this note, we have shown that in a significant area in parameter space we can individually measure the electron and photon couplings of the model to good precision, thereby giving us access to valuable information on the nature of the underlying axion model. Interestingly, the relevant region includes the parameter range suggested by stellar cooling anomalies~\cite{Giannotti:2017hny}. In addition to the couplings, we can also measure the mass of the axion, thereby testing a possible QCD axion nature.

For a first measurement already an energy resolution and threshold of \SI{1}{\kilo \electronvolt} provides good resolving power. In particular in the massive case,
significant improvement and a breaking of degeneracies is possible with an optimized setup that allows for a lower threshold and better energy of the order of  \SI{0.1}{\kilo \electronvolt}.
Therefore, when choosing a detector for IAXO, not only a low background, but also good energy resolution should be taken into account. Promising designs are, for example, micromegas detectors (energy resolution of $\sim$\SI{200}{\electronvolt}) \cite{Krieger:2018nit,Garza:2016nty} and metallic magnetic calorimeters (energy resolution of $\sim$\SI{2}{\electronvolt}) \cite{Kempf2018}. Moreover, since the lower threshold is an important factor in the observed improvement, this motivates studying X-ray optics that would allow IAXO to also efficiently image lower energy photons.

\section*{Note Added}
During the completion of our manuscript we became aware of a paper that also exploits energy resolved detection in axion helioscopes~\cite{Dafni:2018tvj}.
Their work nicely complements our investigation. Their focus is on determining the axion mass, whereas ours is on distinguishing different couplings and we treat the mass measurement only in an exemplified manner.

\section*{Acknowledgements}
The authors would like to thank Loredana Gastaldo for fruitful discussions.
This work was supported by the state of Baden-Württemberg through bwHPC.

\end{document}